\documentclass[journal]{IEEEtran}
\usepackage{amsmath}
\usepackage{amsfonts}
\usepackage{booktabs}
\usepackage{graphicx}
\usepackage{hyperref}
\usepackage{longtable}
\usepackage{multirow}
\usepackage[utf8]{inputenc}
\usepackage[T1]{fontenc}
\usepackage{cite}
\usepackage{fmtcount}

\makeatletter
\newcommand{\aicitenums}{[\@ifundefined{b@aiClaude}{?}{\csname b@aiClaude\endcsname}]--[\@ifundefined{b@aiBGE}{?}{\csname b@aiBGE\endcsname}]}
\makeatother
\newcommand{\aimark}{\textsuperscript{\dag}}
\newcommand{\aifirst}{\begingroup\renewcommand{\thefootnote}{\dag}%
  \footnote{AI-assisted content, in this and every section marked \dag;
  see the Acknowledgment~\aicitenums.}\addtocounter{footnote}{-1}\endgroup}

\newcommand{\pn}[2]{#2}
\newcommand{\lit}[1]{#1}
\DeclareRobustCommand{\pncword}[1]{\ifnum#1<10 \numberstringnum{#1}\else#1\fi}
\DeclareRobustCommand{\Pncword}[1]{\ifnum#1<10 \Numberstringnum{#1}\else#1\fi}
\newcommand{\pnc}[2]{\pncword{#2}}
\newcommand{\Pnc}[2]{\Pncword{#2}}

\begin{document}

\title{Adversarial Process Disturbances at a Deployable Alarm Budget:
Detector Evaluation on a Simulator and a Real Plant}

\author{Benjamin Blakely,~\IEEEmembership{Senior Member,~IEEE}%
\thanks{This work has been submitted to the IEEE for possible publication. Copyright may be transferred without notice, after which this version may no longer be accessible.}
\thanks{B. Blakely is with Argonne National Laboratory, Lemont, IL 60439 USA
(e-mail: bblakely@anl.gov).}%
\thanks{This work was funded by the U.S.\ Department of Energy, Office of
Nuclear Energy, Advanced Reactor Safeguards and Security program, and was
partially supported by the U.S.\ Department of Energy, Office of Science
under DOE contract number DE-AC02-06CH11357. The submitted manuscript
has been created in part by UChicago Argonne, LLC, operator of Argonne National
Laboratory. Argonne, a DOE Office of Science laboratory, is operated
under Contract No.\ DE-AC02-06CH11357. The U.S.\ Government retains for
itself, and others acting on its behalf, a paid-up nonexclusive,
irrevocable worldwide license in said article to reproduce, prepare
derivative works, distribute copies to the public, and perform publicly
and display publicly, by or on behalf of the Government.}
}

\markboth{Preprint submitted to IEEE Transactions on Dependable and Secure Computing}%
{Blakely: Adversarial Process Disturbances at a Deployable Alarm Budget}

\maketitle

\begin{abstract}
Industrial anomaly detection is evaluated largely on simulators, and recent
work reports large drops, with unstable rankings, on real plant data. We ask whether that extends to the
security case. We evaluate \pnc{CountPanelDetectors}{11} detectors across classical, neural,
dependency and causal families on a simulated chemical process and a
sodium thermal-hydraulic facility: different processes sharing no
equipment, but both continuous, at steady state, and perturbed only through
thermal-hydraulic measurements and commands, never composition. \lit{Three} conditions prior work does not
cover: disturbances authored against a threat
model rather than native faults; one construction applied to both
processes; and detection read at a declared false-alarm budget. The difference holds throughout. On the simulator the classical family reaches
$\pn{DetectionClassicalCeilingTpr}{1.00}$ on \pnc{DetectionSimulatorClassicalCeilingCount}{6} of \pnc{CountSimulatorDisturbances}{7} disturbances and the neural family $\pn{DetectionNeuralHighThreshold}{0.89}$ or better on \pnc{DetectionSimulatorNeuralHighCount}{2}, at the second of \pnc{CountTiers}{4} severity tiers and a $\pn{DesignMarkedBudgetLowerPercent}{0.1\%}$ false-positive budget; on plant
telemetry absolute detection rates are not preserved, falling unevenly across detectors and
disturbances, with the neural family lowest. The detector ranking is positively
correlated between the \pnc{CountProcesses}{2} processes at every operating point, but how often a
detector fires does not carry over.
\end{abstract}


\section[Introduction]{Introduction\aifirst}\label{sec:intro}

Cybersecurity anomaly detection for operational technology has largely
inherited its methods, and its evaluation habits, from information technology. That inheritance has
costs which the literature documents rather than disputes: regulatory
qualification burden in safety-critical settings \cite{muhlheimStatusReportRegulatory2023,sumanArtificialIntelligenceNuclear2021},
inference cost on embedded control hardware \cite{chengCheckingBelievingEventAware2021,huangDeepLearningAdvancements2025},
the scarcity of labeled incident data \cite{landauerReviewTimeseriesAnalysis2025,chenClosingSimulationtoRealityGap2026,huangDeepLearningAdvancements2025},
and degradation when a model meets a process other than the one it was fitted
to, which obliges the operator to retrain and requalify it there
\cite{yanComprehensiveSurveyDeep2024,usmaniDeepLearningAnomaly2024,hartungTennesseeEastmanBenchmarking2026,lindemannSurveyAnomalyDetection2021,leeAutoencoderbasedDetectorDistinguishing2024}.

Those costs would pay for themselves if the added capacity bought detection
performance where the detector is deployed. The difficulty is that most of the
evidence that it does is produced somewhere the detector is not deployed.
Simulators and public benchmarks dominate the evaluation literature, and the
field already documents their weaknesses
\cite{choiDeepLearningAnomaly2021,landauerReviewTimeseriesAnalysis2025,kimStudyPerformanceMetrics2022,hartungTennesseeEastmanBenchmarking2026,mudasirRBCADConformalAnomaly2026}, while the relative ranking of methods is
known to shift with the dataset \cite{hartungTennesseeEastmanBenchmarking2026,usmaniDeepLearningAnomaly2024,leeAutoencoderbasedDetectorDistinguishing2024}.

The concern is not a new one: the simulated-to-real gap in industrial anomaly
detection is an established result, benchmarked across \lit{thirty} methods and
\lit{three} processes by Hartung et al.\ and described below
\cite{hartungTennesseeEastmanBenchmarking2026,hartungDeepAnomalyDetection2023}.

What is not established is whether it survives the conditions a security
evaluation imposes. \lit{Three} differences matter. First, the prior result uses
each process's own process faults;
an adversary is not a fault, and a disturbance constructed to stay inside a
plant's admissible envelope (the range each channel is observed to occupy,
with a ceiling above it) is a different object from a valve sticking.
Second, comparing one process's faults against another's leaves the
disturbance and the telemetry varying together; applying one construction to
both removes the disturbance design as a variable, though not the realized
magnitudes, which each process's own envelope sets. Third, performance was
assessed with F1 and area under the precision--recall curve. The curve area
aggregates over thresholds an operator would never select, and F1 is read at a
single threshold rather than at an alarm rate fixed in advance.

We therefore ask a narrower question than the prior work answered: does the
transfer failure persist for adversarial disturbances, under an identical
disturbance construction, at a false-alarm rate a control room could staff?
\footnote{\emph{Adversarial} here means attacker-authored within an admissibility
envelope, not optimized against a deployed detector: an adversary who
queries the detector or maximizes impact beneath its threshold
\cite{urbinaLimitingImpactStealthy2016,erbaAssessingModelfreeAnomaly2022} is
a stronger one than we instantiate, and Section~\ref{sec:limitations} says
so.}
It does. A second question follows from it, and the \lit{two} are separable: whether
the detector ranking that holds under a threshold-free metric still holds at a
budget an operator would actually choose.

A companion technical report covers the corpus construction and the
single-process results in more detail than is reproduced here
\cite{blakelyPhysicsBoundedRobustnessTR2026}.

Our contributions are \lit{three}:

\begin{enumerate}
\item A tiered disturbance framework bounded by a per-channel envelope, whose
bounds come from engineering limits where a process declares them and from its
own telemetry otherwise, portable across process data models, which lets the
same adversary model be instantiated on \pnc{CountProcesses}{2} processes that share no equipment
(Section~\ref{sec:method}).
\item A descriptive comparison, following
\cite{hartungTennesseeEastmanBenchmarking2026}, showing that detection
performance on this simulated process does not reproduce on this real process
for adversarial disturbances, under one disturbance construction applied to
both processes rather than each process's own faults
(Section~\ref{sec:results}).
\item A separation between \lit{two} things often reported together: the detector
\emph{ranking} is positively correlated across the \pnc{CountProcesses}{2} processes at every
operating point examined, while the absolute detection rate is not preserved
(Section~\ref{sec:results}).
\end{enumerate}

\section[Background and Related Work]{Background and Related Work\aimark}\label{sec:background}

\label{sec:mspc}
\emph{Multivariate statistical process monitoring.} Process industries have monitored plants with multivariate statistics for
decades. The procedure was established in the process-control literature by
the early 1990s: project correlated sensor readings onto a low-dimensional
subspace by principal component analysis (PCA) or partial least squares, fit that
subspace to a reference period defining normal operation, and monitor new
observations both within the subspace, by score statistics, and against it, by
squared prediction error, with limits taken from the reference distribution
\cite{krestaMultivariateStatisticalMonitoring1991}. That the approach remains
in use at plant scale, and that fault isolation built on it is still current,
is documented in reviews in the time since
\cite{heStatisticalProcessMonitoring2018,geReviewDatadrivenModeling2017,yoonFaultDiagnosisMultivariate2001}.

Learned detectors are the field's center of gravity in industrial control
systems (ICS) and cyber-physical systems (CPS) alike. Autoencoders, recurrent
and convolutional reconstruction models, graph and transformer architectures
have all been applied to industrial control and cyber-physical telemetry, and
the surveys covering them are numerous enough to have their own
review literature
\cite{huangDeepLearningAdvancements2025,lindemannSurveyAnomalyDetection2021,choiDeepLearningAnomaly2021,zamanzadehdarbanDeepLearningTime2025,yanComprehensiveSurveyDeep2024}.
The case for them is a real one: they require no explicit model of the process,
they capture nonlinear and temporal structure that a static projection cannot,
and they have posted strong results on the standard benchmarks
\cite{peralesgomezInterpretableSemisupervisedSystem2023}.

The question is not whether these methods work, but where the evidence that
they work was collected.

That evidence comes overwhelmingly from a small number of simulated and public
benchmark datasets, and the field says so itself. Reviews of time-series
anomaly detection report that evaluation protocols vary enough to make
published numbers hard to compare, that benchmark datasets carry artifacts
that inflate reported detection, and that fair comparison against classical baselines
is scarcer than the volume of the literature suggests
\cite{rewickiItWorthIt2023,yuDeepLearningClassical2024,zamanzadehdarbanDeepLearningTime2025,mudasirRBCADConformalAnomaly2026}.
Within industrial control specifically, the same observation recurs, and
authors routinely motivate their own contributions by it
\cite{kimComparativeStudyTime2023,giraldoSurveyPhysicsBasedAttack2019,hartungDeepAnomalyDetection2023}.
Where the comparison is made, the classical baseline is not reliably the
weaker: Kravchik and Shabtai report PCA attaining high attack-detection
rates across \lit{three} public ICS datasets given appropriate preprocessing and
feature selection, alongside the lightweight neural detectors that are the
paper's subject \cite{kravchikEfficientCyberAttack2022}. The result here is
a similar finding on a real plant, at a declared operating point, against
disturbances constructed identically for both processes.

The papers cited observe the gap and use it
to justify their own work; they do not jointly establish a quantitative claim
about how large it is.

\label{sec:bg-simreal}
\emph{Simulated versus real evaluation.} The closest prior work to this paper establishes the effect we observe.
Hartung et al.\ benchmark \lit{thirty} anomaly-detection methods (\lit{twenty-six}
deep and \lit{four} classical) through one preprocessing, model-selection and
evaluation pipeline, on the simulated Tennessee Eastman process and on \lit{two}
continuously operated pilot plants, and report average scores on real data far
below those commonly published for the simulator, with no method consistently
best and rankings that depend on both the metric and the process scenario
\cite{hartungTennesseeEastmanBenchmarking2026}. On breadth of methods that result exceeds this one: \lit{thirty} methods against
\pnc{CountPanelDetectors}{11}, and \lit{three} facilities against our \pnc{CountProcesses}{2} processes. It is confined to native process
faults scored by F1 and precision--recall area, neither read at an alarm rate fixed in advance.

Real-plant evidence otherwise rests largely on public testbeds carrying
deliberate attacks, chiefly the Secure Water Treatment (SWaT), Water Distribution (WADI) and
Hardware-in-the-loop Augmented ICS (HAI) testbeds
\cite{mathurSWaTWaterTreatment2016,gohDatasetSupportResearch2017,ahmedWADIWaterDistribution2017,shinHAI10HILbased2020}.
Those are purpose-built and attacked by their own operators; the plant here is
an operating facility whose record was not produced for a security study.

\section[Threat Model]{Threat Model\aimark}\label{sec:threat}

Our evaluation works by perturbation of collected benign data, and the tiers used throughout this paper
are a property of what gets perturbed. We take telemetry recorded while a
process ran normally, apply synthetic disturbances to it, and ask whether a
detector trained on the unperturbed record flags the result. A disturbance is
a named pattern (a sensor drifting, a setpoint nudged the way an operator would, an
oscillation induced in a control loop) intended to represent a prototypical
cyber attack, realized as a modification to specific
channels over a specific interval. Each one is generated at \pnc{CountTiers}{4} severity
tiers, and a tier is defined relative to what that process's own telemetry
and bounds permit rather than as an absolute magnitude. That relative
definition is what lets the same disturbance be generated on \pnc{CountProcesses}{2} processes 
sharing no equipment.

We define the adversary by what they know, and map it onto the severity
ladder. The mapping from knowledge to tier is a modeling assumption, and
admissibility means compliance with the declared per-channel envelope, not
demonstrated physical attainability or process impact. Table~\ref{tab:threat} gives the
mapping. Its third column determines
which rung a detector is the marginal control for and therefore which rung the
results should lead with.

\begin{table}[t]
\caption{Adversary knowledge assumed for each tier, and whether a limit alarm, the simplest control tested, catches it.}
\label{tab:threat}
\centering
\begin{tabular}{@{}llp{2.4cm}@{}}
\toprule
Tier & Adversary knows & Limit alarm \\
\midrule
T1 & the observed telemetry envelope & has nothing to trip on \\
T2 & process physics and channel coupling & has nothing to trip on \\
T3 & physics; willing to sit at the limits & catches disturbances that leave the range \\
T4 & nothing (unconstrained control condition) & catches those, in unbounded mode \\
\bottomrule
\end{tabular}
\end{table}

A tier is a probability distribution over magnitudes, not a range of them.
For a quantity set as a level, which is every disturbance's magnitude, T1
through T3 draw from the same per-channel range and differ only in where
the weight sits inside it, so their draws overlap and a single T1 draw can
exceed a single T3 one. What separates them is how much exposure the
adversary will accept: T1 keeps the disturbance where normal operation
already sits, T2 spreads the impact outside historical observations while
remaining inside the bounds, T3 pushes probability mass to those
bounds, and T4 alone widens the range past them. Sensor drift with dropouts
also draws a rate, and for a rate the shape is fixed and the support expands
from tier to tier. Section~\ref{sec:method} gives both parameterizations.

T1 is by construction hard to
separate from observed variation, since its distribution was fitted to that
variation. T3 and T4 concentrate at or beyond the bounds, where a plant's own
limit alarm is the relevant control for the disturbances that actually leave
the range, though the limit-alarm check below finds that is fewer of them
than the tier label suggests. T2 is the tier whose draws are routinely far from normal
operation and, by construction, never outside the envelope: the simple
limit alarm we test has nothing to trip on, though we do not show that no
cheaper control exists.

\section[Method]{Method\aimark}\label{sec:method}

Both processes are read through one telemetry schema modeled on ISA-95 and
the OPC Unified Architecture (OPC UA) information model, with a per-process adapter converting from the native format on ingest.
Every stage downstream of ingest (perturbation, detection, scoring) sees
only the unified form and contains no process-specific branch. If each process carried its own field names and grouping, a difference in detection could
be attributed to the data model instead of to the data, and there would
be no way to separate the \lit{two} from the outside. Placing the only
process-specific code at ingest makes the rest of the pipeline provably
identical across the \lit{two}.

\label{sec:construction}

\emph{Disturbance construction.} A disturbance function is authored once and applied to each process's recorded telemetry, which is what allows
the same one to run on \pnc{CountProcesses}{2} plants sharing no equipment. It has \lit{three} parts.

The first is a set of \emph{slots}: the parts an instrument plays in that
disturbance, named by function rather than by tag. A flow-oscillation
disturbance, for example, declares a slot for the flow it oscillates, one for
a pressure that should respond, and one for a pump signal that should not.
Slots are filled per process from a registry of instrument roles, written in
the same ISA-95 vocabulary as the telemetry schema: \pnc{DesignPlantRoleCount}{76} roles are
declared for the real plant and \pnc{DesignSimulatorRoleCount}{55} for the simulator, and a
per-process binding file maps each slot to the roles that can fill it. A
disturbance whose slots cannot all be filled does not run there, which is why 
\pnc{CountPlantOnlyDisturbances}{2} of the \pnc{CountPlantDisturbances}{9} are structurally absent on the simulator rather than
merely not evaluated.

\begin{table}[t]
\caption{The \pnc{CountPlantDisturbances}{9} disturbances, \pnc{CountPlantOnlyDisturbances}{2} of which cannot be bound on the
simulator for lack of equipment to fill their slots; FDI is false-data
injection.}
\label{tab:signatures}
\centering
\footnotesize
\begin{tabular}{@{}@{}lp{4.15cm}c@{}}
\toprule
Disturbance & What it does & Sim. \\
\midrule
Coordinated thermocouple & moves several thermocouples together so the spatial pattern stays plausible & \\
Physics violation & drives values outside the channel's observed support; positive control at T3 and T4 & \\
\midrule
Flow oscillation & induces an artificial oscillation in flow channels & \checkmark \\
Operator mimicry & small, infrequent setpoint changes, each plausible as operator action & \checkmark \\
Oscillatory instability & drives a control loop toward unstable behavior & \checkmark \\
Physics consistent & moves coupled channels together so the physical relations still hold & \checkmark \\
Precision trust decay & degrades sensor precision and injects noise over time & \checkmark \\
Replay FDI & replays an earlier window of readings in place of live ones & \checkmark \\
Sensor drift & slow drift combined with intermittent dropouts & \checkmark \\
\bottomrule
\end{tabular}
\end{table}

Table~\ref{tab:signatures} lists the \pnc{CountPlantDisturbances}{9} and what each does. They are not
invented for this study: each is derived from a control flaw identified by a
systems-theoretic process analysis of the facility
\cite{levesonSTPAHandbook2018}, in the security-extended form that treats a
compromised controller or sensor as one more way for a control action to be
unsafe \cite{friedbergSTPASafeSecSafetySecurity2017}. The companion report
gives the flaw-by-flaw mapping
\cite{blakelyPhysicsBoundedRobustnessTR2026}. Between them the \pnc{CountPlantDisturbances}{9} reach
the sensor side and the actuator side, single instruments and coordinated
groups, and they are deliberately uneven in subtlety: one is a positive
control that puts values outside the channels' observed support, while others are built
to survive an operator's inspection.

The second part of the disturbance function is a \emph{transform}. \Pnc{DisturbancePrimitiveCount}{8}
primitives compose the \pnc{CountPlantDisturbances}{9} disturbances: offsets, oscillation, fixed
patterns, replay of an earlier window, injected noise, degraded precision,
dropped updates, and coupling of a change to a physical relationship between
channels. A disturbance is one or more of
these applied to its filled slots.

The third part is the magnitude, which is where the tiers enter. Each slot names a
\emph{magnitude reference}, a channel whose envelope statistics set the scale
for that slot, so a disturbance asks for a deviation sized to the instrument
it is perturbing rather than an absolute value carried over from wherever it
was authored. The admissibility envelopes below give how a tier turns that reference
into a number, and the tiering rules how magnitude, rate and duration
compose.

\label{sec:envelope}

\emph{Admissibility envelopes.} An admissibility envelope records, per channel, the range that channel has been
observed to occupy and a ceiling above it. Bounds come from a declared source
where one exists and from the channel's own observed statistics otherwise. The
envelope is the only place a disturbance learns how far it may move a given
signal, and it is what turns a tier into a number.

Write $\ell$ and $S$ for the lower bound and width of a channel's admissible
range, and $\mu$ and $\sigma$ for its mean and standard deviation as
fractions of that range. Every level draw takes the form:
\begin{equation}
x = \ell + S\,b, \qquad b \sim \mathrm{Beta}(\alpha, \beta),
\label{eq:tierdraw}
\end{equation}
and, for a level, the tiers differ only in how they set $(\alpha, \beta)$. T1 and T2 match
moments to a target variance $\sigma_\star^2$:
\begin{equation}
\alpha = \mu\nu, \quad \beta = (1-\mu)\nu, \quad
\nu = \max\!\left(\frac{\mu(1-\mu)}{\sigma_\star^2} - 1,\, \nu_0\right),
\label{eq:moments}
\end{equation}
taking $\sigma_\star^2 = \sigma^2$ for T1 and the channel-independent
$\sigma_\star^2 = (1/6)^2$ for T2. The concentration $\nu = \alpha + \beta$ is what the
target buys: the variance is $\mu(1-\mu)/(\nu+1)$, so tiers loosen as $\nu$
falls.
The floor $\nu_0$ ($\pn{TierBetaMomentFloorTierOne}{2}$ for T1, $\pn{TierBetaMomentFloorTierTwo}{1}$ for T2) prevents a degenerate draw.
T3 and T4 set the shape directly instead, at
$\alpha = \max(\lit{0.8}\mu,\, \lit{0.1})$ and $\beta = \max(\lit{0.8}(1-\mu),\, \lit{0.1})$. Both
parameters fall below one on any channel, inverting the curve so mass collects
at the \lit{two} ends rather than the middle. They differ only in range: T4 widens
it to $[\ell - 10\sigma S,\; \ell + S + 10\sigma S]$ and recomputes $\mu$
within it, which is the only way a draw can leave the admissible bounds.
Table~\ref{tab:tiers} gives the resulting parameters for one channel.

\begin{table}[t]
\caption{The tiers as inputs to (\ref{eq:tierdraw})--(\ref{eq:moments}), for
one simulator channel with mean at $\pn{TierBetaPurgeValveMean}{0.551}$ of its admissible range. $\alpha$ and
$\beta$ are that channel's, not constants of the tier.}
\label{tab:tiers}
\centering
\begin{tabular}{@{}llrr@{}}
\toprule
Tier & Target variance $\sigma_\star^2$ & $\alpha$ & $\beta$ \\
\midrule
T1 & $\sigma^2$, the channel's own & $\pn{TierBetaPurgeValveTierOneAlpha}{53.63}$ & $\pn{TierBetaPurgeValveTierOneBeta}{43.73}$ \\
T2 & $(1/6)^2$ & $\pn{TierBetaPurgeValveTierTwoAlpha}{4.36}$ & $\pn{TierBetaPurgeValveTierTwoBeta}{3.55}$ \\
T3 & shape set directly & $\pn{TierBetaPurgeValveTierThreeAlpha}{0.44}$ & $\pn{TierBetaPurgeValveTierThreeBeta}{0.36}$ \\
T4 & as T3, range widened & $\pn{TierBetaPurgeValveTierFourAlpha}{0.42}$ & $\pn{TierBetaPurgeValveTierFourBeta}{0.38}$ \\
\bottomrule
\end{tabular}
\end{table}

Because $\mu$ enters both parameters, a tier fixes a rule rather than a shape:
the curve leans toward the end each channel normally operates near, so the
same tier means a different absolute magnitude on every one. That is what
makes a single disturbance construction comparable across processes sharing
no equipment.

A rate, which only sensor drift with dropouts draws, comes from
a ladder built the opposite way around: the shape stays a symmetric bell
($\alpha = \beta = 2$) and the support expands, reaching at T1 through T4 the
channel's observed $99$th percentile, its observed maximum, the
ceiling, and that ceiling plus \pnc{TierRateCeilingStandardDeviations}{10} rate standard deviations.

Both the range and that ceiling are therefore per channel, and they have to
be: a global constant gives every channel the same admissible interval, which
is defensible only where every channel measures the same quantity in the same
units. The simulator is not such a case -- its per-channel rate ceilings span
a factor of \pn{SimulatorBoundsRateCeilingSpanFactor}{5,201}, from $\pn{SimulatorBoundsRateCeilingMinimum}{0.053}$ to $\pn{SimulatorBoundsRateCeilingMaximum}{275.67}$ per second, across kscmh/s
(\lit{thousand} standard cubic meters per hour, per second), kPa/s, mol\%/s and \%/s -- and under one ceiling T3 and T4 would key off the
same scalar and stop differing. An envelope that hard-codes one process's
answer is silently wrong on the next rather than visibly absent.
Section~\ref{sec:limitations} states how much of each process's envelope is
independently sourced and how much is empirical.

\label{sec:tiered}

\emph{Tiering and clamping.} A disturbance is realized by drawing a magnitude from that channel's
tier distribution and applying the transform. Timing is not derived from that
magnitude: each primitive carries its own base duration -- an outage length, a
replay window, a toggle interval -- scaled by a per-tier severity factor
($\pn{DesignTierOneSeverityFactor}{0.15}$, $\pn{DesignTierTwoSeverityFactor}{0.55}$, $\pn{DesignTierThreeSeverityFactor}{0.95}$ and $\pn{DesignTierFourSeverityFactor}{1.30}$ from T1 to T4),
since a blackout and a drift share no unit in which one duration rule could be
written.

Once a magnitude is drawn, the disturbance can meet the envelope in \lit{two} ways.
In \emph{bounded} clamping mode every perturbed value is checked against its
channel's admissible support. At T4 a value outside that support is clamped
back inside, and that clamp is the contrast the mode exists to create. At T1,
T2 and T3 a value outside it means the tier contract is broken rather than that
something needs correcting, so generation halts, the exception being a channel
whose baseline leaves no admissible room for the draw at all, which is clamped
and counted separately. In \emph{unbounded} mode none of this runs.

Every detector result reported here is bounded, the limit-alarm baseline of
Section~\ref{sec:results} excepted for the reason given there, because
bounded is the mode in which
the tier definitions are enforced rather than merely intended: a T1 corpus
generated unbounded carries no guarantee of admissibility, the one property
T1 is named for.

The choice costs something at T4, the only tier whose draws are meant to leave
the admissible range. The plant was run both ways. Comparing the \pnc{DesignPlantClampingModeCount}{2} arms
there, the clamp alters, averaged without weighting over the \pnc{CountPlantDisturbances}{9} disturbances, $\pn{ClampTierFourMeanRate}{16.1\%}$ of the
rows at which the disturbance is active in both arms at T4, $\pn{ClampTierThreeMeanRate}{4.9\%}$ at T3 and none at T1 and T2, so T4 keeps most of its excursion rather than collapsing onto
T3, but it is a softer positive control than the unclamped definition
describes. Section~\ref{sec:results} reports a paired comparison that finds no
significant difference between the modes, a weak test at \pnc{ClampModePairedSeedCount}{5} seeds.

The simulator was run bounded only, so the unclamped arm exists for the plant
alone. The cross-process comparison is bounded on both sides.

\label{sec:detectors}

\emph{Detector suite and fairness protocol.} \Pnc{CountPanelDetectors}{11} detectors span \pnc{CountFamilies}{4} families. The classical family is Hotelling's
$T^2$, PCA reconstruction, isolation forest and change-point detection. The
neural family is an autoencoder and a long short-term memory (LSTM)
forecaster. The dependency family
is a Bayesian network and a correlation and Granger-based dependency-violation
detector. The causal family is DYNOTEARS \cite{pamfilDYNOTEARSStructureLearning2020},
NTS-NOTEARS (nonlinear time-series NOTEARS, after NOTEARS
\cite{zhengDAGsNOTEARS2018}) and PCMCI+ (the PC algorithm with momentary
conditional independence tests, extended to contemporaneous links)
\cite{rungeDiscoveringContemporaneousLagged2020}, a constraint-based
conditional-independence method.
The supplementary appendix gives each configuration.

All \pnc{CountPanelDetectors}{11} see the same preprocessed corpus and the same splits, and none is
tuned per disturbance, per tier or per process. Each trains only on
unperturbed data from its own process. All \pnc{CountPanelDetectors}{11} run at \pnc{DesignPlantSeedsPerDetectorCount}{5} seeds on both
processes. \Pnc{SeedSpreadIdenticalAucFourDecimalsDetectorCount}{6} of them return the same area under the ROC curve (AUC, Section~\ref{sec:protocol}) at every seed to \lit{four} decimal
places; the spread of the rest is reported in Section~\ref{sec:robustness}.

One asymmetry in the inputs survives. \Pnc{DesignChannelCappedDetectorCount}{4} of the \pnc{CountPanelDetectors}{11} detectors see fewer
channels than the rest: the \pnc{DesignCausalChannelCappedDetectorCount}{3} causal detectors at $\pn{DesignCausalChannelCapMinimum}{20}$ to $\pn{DesignCausalChannelCapMaximum}{30}$ of that
process's channels, and the Bayesian network at $\pn{DesignBayesianNetworkChannelCap}{25}$. The causal detectors
are capped because the computational cost of learning a graph grows steeply in the number of
channels, steeply enough that full width was not affordable for all of them.

Which channels they see matters: a capped detector that cannot observe the
disturbed instrument scores zero by construction. Selection
stratifies by equipment group and assigns round-robin, because selecting the
highest-variance channels admits no commandable channel on this plant.

PCMCI+ did not complete even a $20$-channel fit inside a \pnc{DesignCausalScalingPcmciPlusCapMinutes}{30}-minute cap,
against \pnc{DesignCausalScalingDynotearsSecondsRounded}{50} seconds for DYNOTEARS at the same width. The causal detectors
and the Bayesian network therefore run at a reduced channel budget whose effect on
detection is unmeasured.

\label{sec:protocol}
\emph{Evaluation protocol.} Detection is reported as true-positive rate (TPR) at a declared
false-positive rate (FPR) budget, not as area under the receiver operating
characteristic curve (ROC, and AUC for the area under it). Both rates are
per sample throughout: a TPR is the fraction of perturbed instants flagged,
which is not the same as the fraction of disturbances noticed. Both the budget and the decision to report at it were fixed before any
result was seen.

This is not a departure from process-monitoring practice so much as a return
to it. The founding reference in multivariate statistical
process monitoring already frames the design problem
as a tradeoff against alarm rate: too narrow a reference set and the
procedure alarms too often, too wide and its sensitivity to abnormal operation
is impaired \cite{krestaMultivariateStatisticalMonitoring1991}. Reporting at a
declared budget makes that tradeoff explicit rather than integrating over it.

Operating points are read off the ROC as the staircase it is. The
false-positive rate jumps each time the threshold crosses another normal
sample's score, so the points a deployed threshold can occupy are the corners
and nothing between them; we report the true-positive rate at the last corner
inside the budget. Reading the sloped line a plotting routine draws between
\lit{two} corners instead reports a detector no threshold produces, and does so
exactly where a tight budget falls, since that is where the steps are widest.
Computing AUC over those same connected corners is not inconsistent with
this. An operating point is a configuration someone deploys; AUC is a summary
over every threshold at once, equal to the probability that a perturbed
sample outscores a normal one with ties counted as half, and the connecting
segments are how that calculation handles ties.

The budget is set from the sampling cadence rather than chosen for
convenience, and both processes share it: the plant historian is resampled to
a $\pn{DesignSamplingGridSeconds}{30}$-second grid and the simulator's $\pn{DesignSimulatorNativeSampleRateHertz}{1}$\,Hz output downsampled to the same,
so for a detector that decides on every sample a given false-positive rate is the same alarm load on each. A detector
deciding once per \pnc{DesignSamplingGridSeconds}{30}-second sample makes about
\pn{DesignDecisionsPerDay}{2,880} decisions a day, so a false-positive rate of $\pn{DesignMarkedBudgetLowerPercent}{0.1\%}$
costs roughly \pnc{DesignFalseAlarmsPerDayLowerBudgetRounded}{3} false alarms a day and $\pn{DesignMarkedBudgetUpperPercent}{1\%}$ costs roughly \pnc{DesignFalseAlarmsPerDayUpperBudgetRounded}{29}.
We report at $\pn{DesignMarkedBudgetLowerPercent}{0.1\%}$ throughout and show $\pn{DesignMarkedBudgetUpperPercent}{1\%}$ alongside it where the
difference matters, so that the tenfold difference in alarm load stays visible
instead of being integrated away. \Pnc{DesignFalseAlarmsPerDayLowerBudgetRounded}{3} a day is a small enough load to argue
about; what a particular control room will tolerate is a site decision, and
what the cadence supplies is the conversion that lets a site make it. The
budget is a ceiling on the alarm rate rather than a count of alarms. Some detectors score every sample; others score once per sliding window
of samples, and on the same cell the LSTM emits about one score for every
\pnc{DesignLstmScoresPerPerSampleScoreRatioRounded}{20} a per-sample detector emits. A detector that decides less often raises
fewer alarms at the same false-positive rate, so the budget fixes where each
detector sits statistically, not how many interruptions it causes. Those
cells also rest on correspondingly fewer independent measurements.

No adequacy threshold is applied to the detection rate. The budget is
declared because it follows from the cadence, which is fixed before any result
is seen; any line separating "adequate" from "inadequate"
detection would instead be chosen after seeing the numbers.

The
evaluated window lies inside the training period rather
than after it, so these are not measurements of temporal generalization. The
estimand is a paired counterfactual: a perturbed window scored against the
matched clean twin of the same period, by a model trained on unperturbed data
that includes that period. That is the right comparison for the question asked
here, which is whether a disturbance is distinguishable from the process's own
normal behavior, and it is the wrong one for asking whether a detector
transfers forward in time. It does assume the paired period's presence in
training is not itself what produces the score, and because a high-capacity
model fits its training data more tightly than a low-capacity one, that
assumption bears unequally on the families being compared. The bias
this introduces across families and across the \pnc{CountProcesses}{2} processes is unmeasured;
Section~\ref{sec:limitations} states it as a limitation.

\section[Experimental Design]{Experimental Design\aimark}\label{sec:design}

The real facility is a sodium thermal-hydraulic test loop with $\pn{DesignPlantChannelCount}{143}$ channels;
the simulated one is the Tennessee Eastman process (TEP), which is public,
with $\pn{DesignSimulatorChannelCount}{53}$. \Pnc{CountPlantDisturbances}{9} disturbances are scored on the plant in \pnc{DesignPlantClampingModeCount}{2} clamping modes and
\pnc{CountSimulatorDisturbances}{7} on the simulator in one, giving \pn{DesignPlantScoredCellCount}{3,960} and \pn{DesignSimulatorScoredCellCount}{1,540} scored cells; the
cross-process comparison is the \pnc{CountSharedDisturbances}{7}-disturbance bounded intersection.
Simulation status and process identity are confounded here, so the design
cannot attribute the difference in detection to simulation as such.

Pairing a chemical process with a sodium loop invites the objection that the
\lit{two} are not comparable at all. What is compared is narrower than the
processes. Across the \pnc{CountSharedDisturbances}{7} shared disturbances, every channel perturbed on
either side is a thermal-hydraulic measurement (temperature, pressure,
flow, level or damper position) or a command driving one (a valve, pump,
fan, heater or setpoint), and not one is a composition measurement: the simulator's \pnc{DesignSimulatorCompositionChannelCount}{19} feed, purge-gas and
product analyzers, which are what make it distinctively a chemical process,
are never touched. Its perturbed channels include the reactor and condenser
cooling-water loops. That is a flow driven to remove heat, read through
temperature and flow, which is functionally the element the plant consists
of. Only the directly perturbed channels are thermal-hydraulic: the
detectors without a channel cap see the whole of each process's telemetry, including the simulator's
composition channels, and the \pnc{DesignChannelCappedDetectorCount}{4} capped ones a subset of it (Section~\ref{sec:detectors}).

What the argument uses from the real facility is its channel count, its
instrument families and the disturbances its topology admits. The plant
description, instrumentation inventory and pipeline schematic are in the
companion technical report \cite{blakelyPhysicsBoundedRobustnessTR2026}.

\Pnc{CountPlantOnlyDisturbances}{2} of the \pnc{CountPlantDisturbances}{9} disturbances cannot be expressed on the simulated facility
at all: it has no thermocouple gradients for the one, and none of the cold-trap
auxiliaries the other targets. They are
structurally absent, not unrun, and we report that gap in the panel instead of
working around it.

Every run is executed in a pinned container, seeded, and recorded with the
commit that produced it. Each stage writes under a version identifier hashed
from its own version and configuration, and each scored cell records
fingerprints of the input it read and the corpus that input derived from, so
a cell reusing a corpus built under a superseded envelope is visible rather
than producing a plausible number nothing would flag. That is what lets us
assert every cell behind Section~\ref{sec:results} read the corpus for the
process currently selected.

The plant record is segmented into operating regimes before anything else
runs, to establish the study's scope by measurement rather than assuming it.
The procedure fixes no thresholds by hand: it segments the record by
changepoint detection on per-channel z-score, volatility and rate of change,
clusters the segments with the cluster count chosen by Bayesian information
criterion, and labels each cluster by its nearest archetype among steady
state, transient, startup, shutdown and trouble, or \emph{unknown} where none
wins by a margin. A segment containing an operator setpoint change is labeled
transient regardless of where it clusters.

Of \pn{DesignRegimeSegmentCount}{2,796} labeled segments, $\pn{DesignRegimeSteadyStateSegmentShare}{96.4\%}$ are steady state and $\pn{DesignRegimeTransientSegmentShare}{3.4\%}$
transient, with no startup, shutdown or trouble segments at all. These are
shares of segments, which differ in length, not of operating time. The
results therefore describe a process labeled steady state in nearly every
segment, with the consequences taken up in
Section~\ref{sec:discussion}.

The labels describe the record; they do not select from it. Detectors train on
the full unperturbed history and the negative class is sampled from it without
regard to regime. Replay is the one disturbance that perturbs only some
instants of an attacked hour, and for it the negative class also includes
that hour's unperturbed instants.
Admitting every regime is deliberate rather than an omission: a detector
trained only on the most common regime learns to flag the rarer ones as
anomalous, and would then appear to detect any disturbance that happened to
land in a transient window. Admitting every regime costs some sharpness in the
baseline and removes that confound.

\section[Results]{Results\aimark}\label{sec:results}

\begin{table*}[t]
\caption{Best detection within each family at tier~2 and a $\pn{BudgetOnePermille}{0.001}$
false-positive budget (Cls classical, Neu neural, Dep dependency, Cau causal;
FDI false-data injection). The families have \pnc{CountClassicalFamilyDetectors}{4}, \pnc{CountNeuralFamilyDetectors}{2}, \pnc{CountDependencyFamilyDetectors}{2} and \pnc{CountCausalFamilyRankedDetectors}{2} members
once PCMCI+ is excluded for too few distinct scores, so a best-of favors the
larger; $\ddag$ marks a median that moves by more than $\pn{DetectionSeedMarkThreshold}{0.06}$ across seeds.}
\label{tab:detection}
\centering
\begin{tabular}{@{}lrrrrcrrrr@{}}
\toprule
& \multicolumn{4}{c}{Real plant} & & \multicolumn{4}{c}{TEP simulator} \\
\cmidrule{2-5}\cmidrule{7-10}
Disturbance & Cls & Neu & Dep & Cau & & Cls & Neu & Dep & Cau \\
\midrule
Coordinated thermocouple & \pn{DetectionPlantCoordinatedThermocoupleManipulationClassicalTpr}{1.00} & \pn{DetectionPlantCoordinatedThermocoupleManipulationNeuralTpr}{0.12} & \pn{DetectionPlantCoordinatedThermocoupleManipulationDependencyTpr}{0.02} & \pn{DetectionPlantCoordinatedThermocoupleManipulationCausalTpr}{1.00} & & \pn{DetectionSimulatorCoordinatedThermocoupleManipulationClassicalTpr}{---} & \pn{DetectionSimulatorCoordinatedThermocoupleManipulationNeuralTpr}{---} & \pn{DetectionSimulatorCoordinatedThermocoupleManipulationDependencyTpr}{---} & \pn{DetectionSimulatorCoordinatedThermocoupleManipulationCausalTpr}{---} \\
Flow oscillation         & \pn{DetectionPlantFlowOscillationClassicalTpr}{0.06} & \pn{DetectionPlantFlowOscillationNeuralTpr}{0.01} & \pn{DetectionPlantFlowOscillationDependencyTpr}{0.00} & \pn{DetectionPlantFlowOscillationCausalTpr}{0.04} & & \pn{DetectionSimulatorFlowOscillationClassicalTpr}{1.00} & \pn{DetectionSimulatorFlowOscillationNeuralTpr}{0.34} & \pn{DetectionSimulatorFlowOscillationDependencyTpr}{0.84} & \pn{DetectionSimulatorFlowOscillationCausalTpr}{1.00} \\
Operator mimicry         & \pn{DetectionPlantOperatorMimicryClassicalTpr}{1.00} & \pn{DetectionPlantOperatorMimicryNeuralTpr}{0.12} & \pn{DetectionPlantOperatorMimicryDependencyTpr}{0.02} & \pn{DetectionPlantOperatorMimicryCausalTpr}{0.65}\pn{DetectionPlantOperatorMimicryCausalSeedMark}{\textsuperscript{\ddag}} & & \pn{DetectionSimulatorOperatorMimicryClassicalTpr}{1.00} & \pn{DetectionSimulatorOperatorMimicryNeuralTpr}{0.01} & \pn{DetectionSimulatorOperatorMimicryDependencyTpr}{0.01} & \pn{DetectionSimulatorOperatorMimicryCausalTpr}{1.00} \\
Oscillatory instability  & \pn{DetectionPlantOscillatoryInstabilityClassicalTpr}{1.00} & \pn{DetectionPlantOscillatoryInstabilityNeuralTpr}{0.01} & \pn{DetectionPlantOscillatoryInstabilityDependencyTpr}{0.02} & \pn{DetectionPlantOscillatoryInstabilityCausalTpr}{1.00}\pn{DetectionPlantOscillatoryInstabilityCausalSeedMark}{\textsuperscript{\ddag}} & & \pn{DetectionSimulatorOscillatoryInstabilityClassicalTpr}{1.00} & \pn{DetectionSimulatorOscillatoryInstabilityNeuralTpr}{0.90} & \pn{DetectionSimulatorOscillatoryInstabilityDependencyTpr}{0.95} & \pn{DetectionSimulatorOscillatoryInstabilityCausalTpr}{1.00} \\
Physics consistent       & \pn{DetectionPlantPhysicsConsistentManipulationClassicalTpr}{1.00} & \pn{DetectionPlantPhysicsConsistentManipulationNeuralTpr}{0.02} & \pn{DetectionPlantPhysicsConsistentManipulationDependencyTpr}{0.02} & \pn{DetectionPlantPhysicsConsistentManipulationCausalTpr}{0.09}\pn{DetectionPlantPhysicsConsistentManipulationCausalSeedMark}{\textsuperscript{\ddag}} & & \pn{DetectionSimulatorPhysicsConsistentManipulationClassicalTpr}{1.00} & \pn{DetectionSimulatorPhysicsConsistentManipulationNeuralTpr}{0.41}\pn{DetectionSimulatorPhysicsConsistentManipulationNeuralSeedMark}{\textsuperscript{\ddag}} & \pn{DetectionSimulatorPhysicsConsistentManipulationDependencyTpr}{1.00} & \pn{DetectionSimulatorPhysicsConsistentManipulationCausalTpr}{0.83}\pn{DetectionSimulatorPhysicsConsistentManipulationCausalSeedMark}{\textsuperscript{\ddag}} \\
Physics violation        & \pn{DetectionPlantPhysicsViolationClassicalTpr}{1.00} & \pn{DetectionPlantPhysicsViolationNeuralTpr}{0.10}\pn{DetectionPlantPhysicsViolationNeuralSeedMark}{} & \pn{DetectionPlantPhysicsViolationDependencyTpr}{0.02} & \pn{DetectionPlantPhysicsViolationCausalTpr}{0.00} & & \pn{DetectionSimulatorPhysicsViolationClassicalTpr}{---} & \pn{DetectionSimulatorPhysicsViolationNeuralTpr}{---} & \pn{DetectionSimulatorPhysicsViolationDependencyTpr}{---} & \pn{DetectionSimulatorPhysicsViolationCausalTpr}{---} \\
Precision trust decay    & \pn{DetectionPlantPrecisionTrustDecayClassicalTpr}{1.00} & \pn{DetectionPlantPrecisionTrustDecayNeuralTpr}{0.00} & \pn{DetectionPlantPrecisionTrustDecayDependencyTpr}{1.00} & \pn{DetectionPlantPrecisionTrustDecayCausalTpr}{0.07} & & \pn{DetectionSimulatorPrecisionTrustDecayClassicalTpr}{1.00} & \pn{DetectionSimulatorPrecisionTrustDecayNeuralTpr}{0.79} & \pn{DetectionSimulatorPrecisionTrustDecayDependencyTpr}{1.00} & \pn{DetectionSimulatorPrecisionTrustDecayCausalTpr}{1.00} \\
Replay FDI               & \pn{DetectionPlantReplayBasedFdiClassicalTpr}{0.00} & \pn{DetectionPlantReplayBasedFdiNeuralTpr}{0.00} & \pn{DetectionPlantReplayBasedFdiDependencyTpr}{0.00} & \pn{DetectionPlantReplayBasedFdiCausalTpr}{0.00} & & \pn{DetectionSimulatorReplayBasedFdiClassicalTpr}{0.00} & \pn{DetectionSimulatorReplayBasedFdiNeuralTpr}{0.00} & \pn{DetectionSimulatorReplayBasedFdiDependencyTpr}{0.00} & \pn{DetectionSimulatorReplayBasedFdiCausalTpr}{0.00} \\
Sensor drift             & \pn{DetectionPlantSensorDriftWithDropoutsClassicalTpr}{1.00} & \pn{DetectionPlantSensorDriftWithDropoutsNeuralTpr}{0.10} & \pn{DetectionPlantSensorDriftWithDropoutsDependencyTpr}{0.02} & \pn{DetectionPlantSensorDriftWithDropoutsCausalTpr}{0.17}\pn{DetectionPlantSensorDriftWithDropoutsCausalSeedMark}{\textsuperscript{\ddag}} & & \pn{DetectionSimulatorSensorDriftWithDropoutsClassicalTpr}{1.00} & \pn{DetectionSimulatorSensorDriftWithDropoutsNeuralTpr}{1.00} & \pn{DetectionSimulatorSensorDriftWithDropoutsDependencyTpr}{1.00} & \pn{DetectionSimulatorSensorDriftWithDropoutsCausalTpr}{1.00} \\
\bottomrule
\end{tabular}
\end{table*}

\label{sec:sim}
\emph{The panel on a simulated process.} Read the right-hand block of Table~\ref{tab:detection} alone and the panel
detects most disturbances. At an alarm rate a control room could staff, the classical
family reaches $\pn{DetectionClassicalCeilingTpr}{1.00}$ on \pnc{DetectionSimulatorClassicalCeilingCount}{6} of the \pnc{CountSimulatorDisturbances}{7} disturbances and the neural
family reaches $\pn{DetectionNeuralHighThreshold}{0.89}$ or better on \pnc{DetectionSimulatorNeuralHighCount}{2}. This is the kind of result conventional evaluation practice
produces, now under adversarial disturbances rather than
the native process faults of the prior benchmark
\cite{hartungTennesseeEastmanBenchmarking2026}.

\label{sec:real}
\emph{The same panel on real plant telemetry.} The left-hand block is the same pipeline, the same disturbance construction,
the same tiers and the same budget. What differs is the
telemetry and the per-channel envelopes drawn from it, the control the prior
comparison could not have, since each process there contributes its own
faults, so disturbance and telemetry vary together. Absolute detection rates are not preserved, and
the neural family again trails the classical one, exceeding a true-positive
rate of $\pn{DetectionNeuralFloorThreshold}{0.10}$ on \pnc{DetectionPlantNeuralAboveFloorCount}{4} disturbances out of \pnc{CountPlantDisturbances}{9}. The direction is consistent with
\cite{hartungTennesseeEastmanBenchmarking2026} on a different pair of
processes, a different disturbance class and a different metric.

\begin{figure*}[t]
\centering
\includegraphics[width=\textwidth]{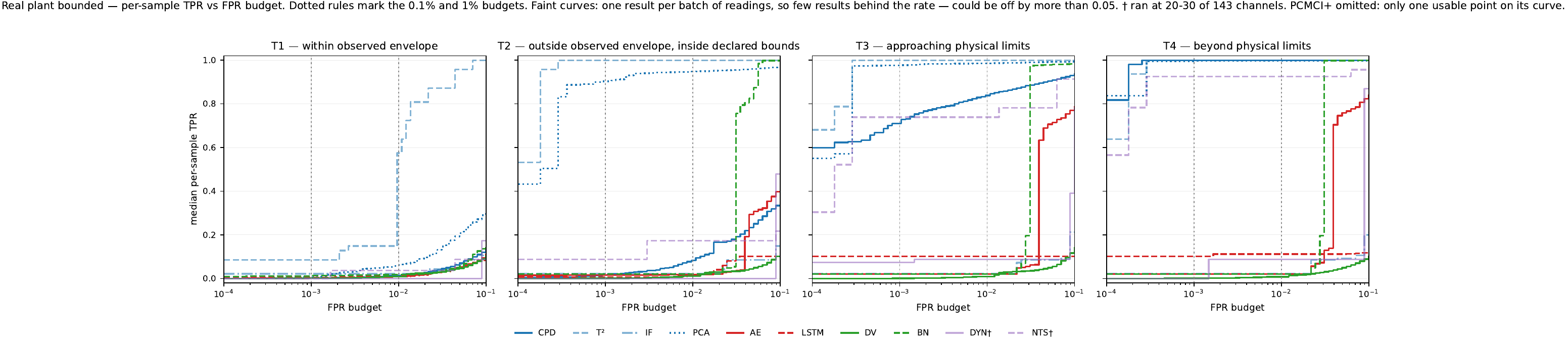}\\[2ex]
\includegraphics[width=\textwidth]{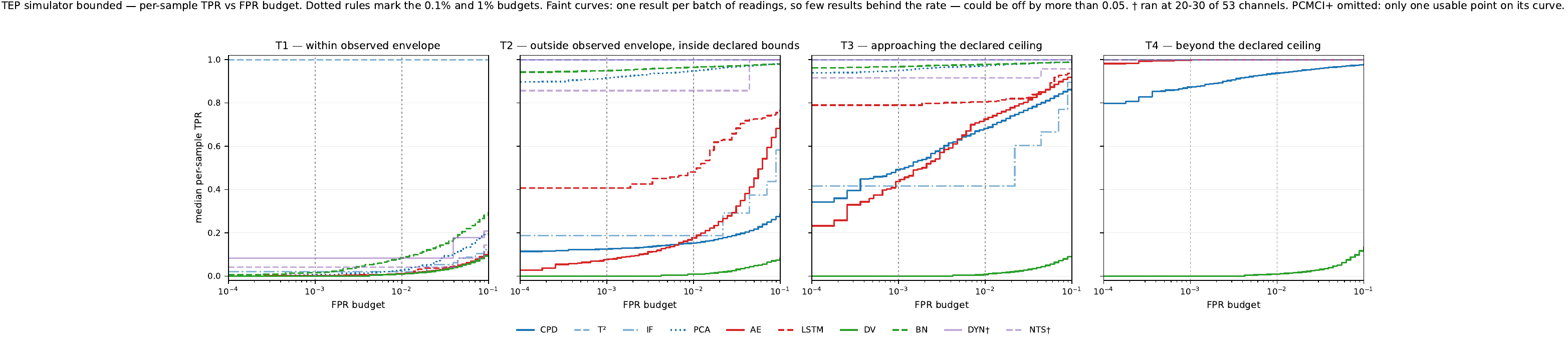}
\caption{Per-sample detection against false-alarm budget, plant above and
simulator below, by tier; rules mark the $\pn{DesignMarkedBudgetLowerPercent}{0.1\%}$ and $\pn{DesignMarkedBudgetUpperPercent}{1\%}$ budgets. CPD
change-point detection, IF isolation forest, AE autoencoder, DV dependency
violation, BN Bayesian network, DYN DYNOTEARS, NTS NTS-NOTEARS.}
\label{fig:budget}
\end{figure*}

Figure~\ref{fig:budget} shows the full ladder at every budget, with curves
drawn faint where few independent measurements stand behind them. Detection results are
reported at tier~2 throughout, for the reason Section~\ref{sec:threat} gives,
except the rank analysis below, which pools all \pnc{CountTiers}{4} tiers. Measured by mean
AUC over disturbances, detection on the plant rises from tier~1 to tier~4 for
\pn{TierLadderRisingDetectorCount}{9} of the \pnc{CountPanelDetectors}{11} detectors; \pn{TierLadderMonotoneRisingDetectorCount}{8} of those rise at every step, and the exception,
DYNOTEARS, dips by $\pn{TierLadderDipSize}{0.002}$ at tier~3. The
remaining \pn{TierLadderNotRisingDetectorCount}{2} are flat rather than inverted. By mean AUC the ladder therefore largely behaves as constructed. It is not
uniformly monotone at the operating points: over the \pnc{CountBudgets}{5} budgets and \pnc{CountPanelDetectors}{11}
detectors, $\pn{DetectionLadderPlantReversalCount}{6}$ of the $\pn{DetectionLadderPlantCurveCount}{55}$ detector-budget curves reverse somewhere on the
plant and $\pn{DetectionLadderSimulatorReversalCount}{5}$ of $\pn{DetectionLadderSimulatorCurveCount}{55}$ on the simulator. Monotonicity is therefore a property of the
ladder overall rather than a guarantee for every detector at every budget.

\begin{figure*}[t]
\centering
\includegraphics[width=0.49\textwidth]{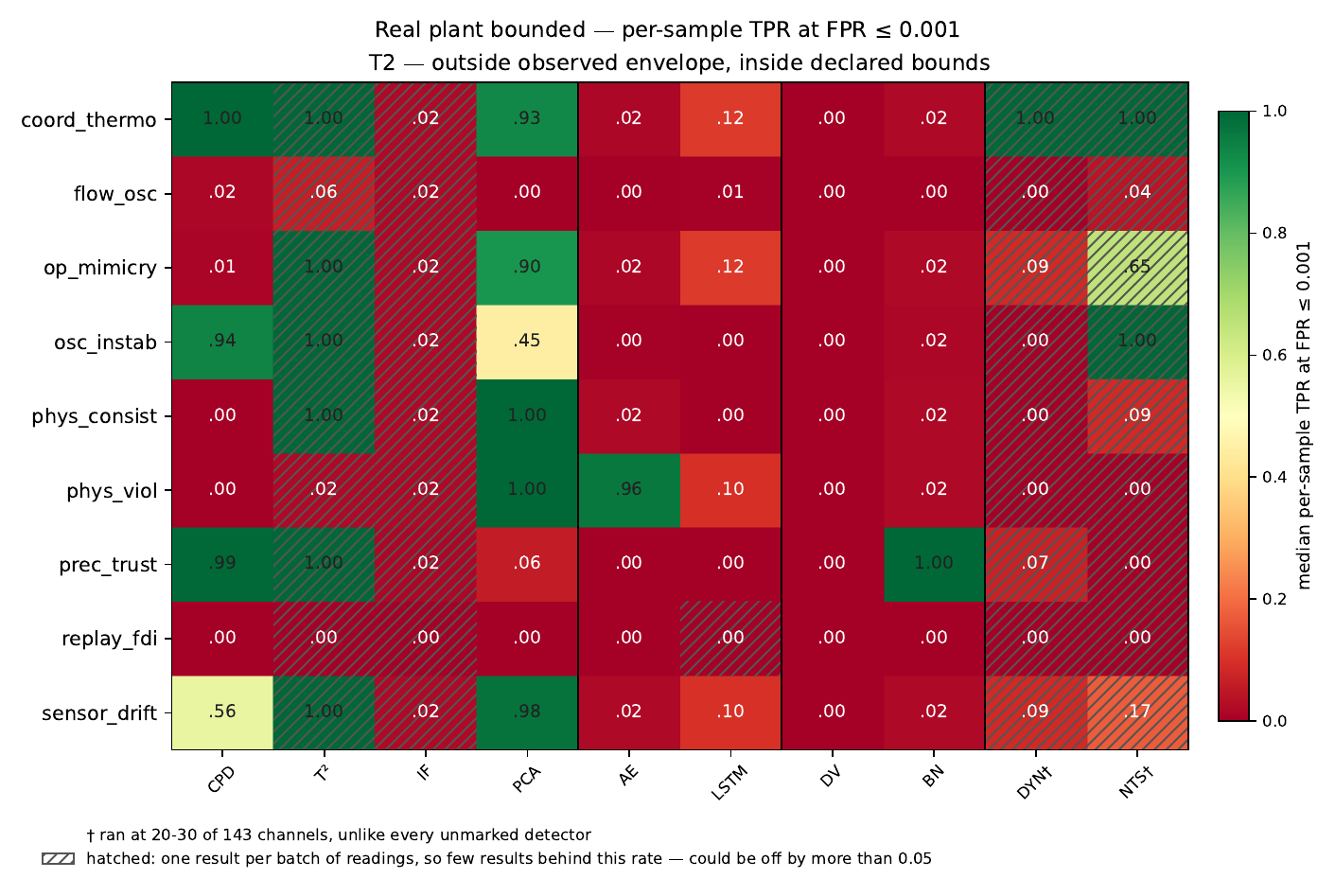}
\hfill
\includegraphics[width=0.49\textwidth]{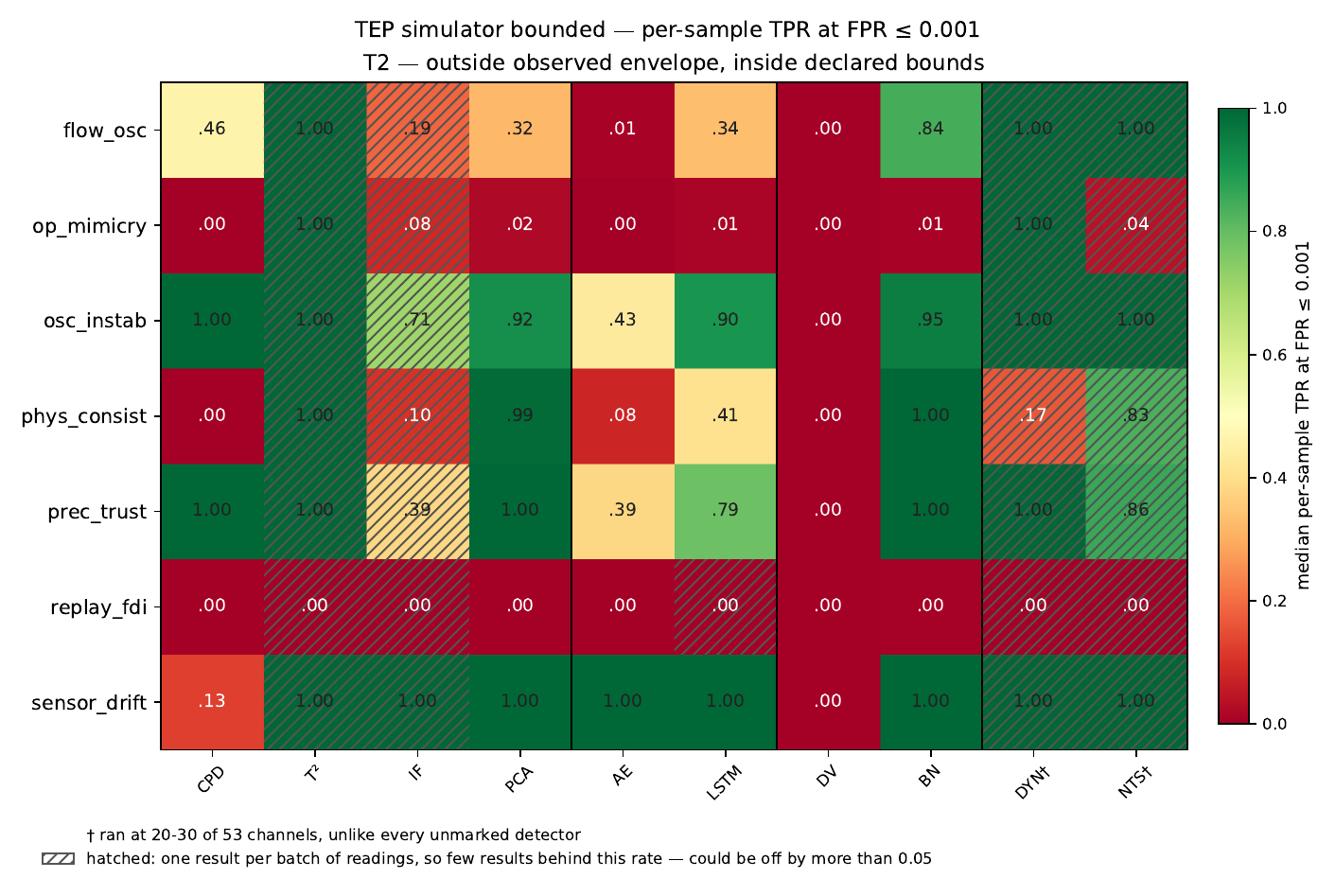}
\caption{Per-detector, per-sample detection at tier~2 and a $\pn{BudgetOnePermille}{0.001}$ budget,
plant left and simulator right, grouped by family; dark means few perturbed
instants flagged, not that the disturbance passed unnoticed (abbreviations as
in Fig.~\ref{fig:budget}).}
\label{fig:heatmap}
\end{figure*}

\emph{A detector can only see what it measures.} Detection also depends on
what each detector measures. Change-point detection works from how fast each channel is
changing: it scores a weighted sum of the first difference, the second
difference and a rolling variance (supplementary Table~A.I). \lit{Two} of our
disturbances leave that quantity unchanged or reduce it, and the detector misses both.

Figures here are AUC rather than the true-positive rate used elsewhere: what
is at issue is whether the detector separates the classes at all, and a
detector at chance has no operating point worth reading.

The first is a constant offset. Shifting a channel by a fixed amount moves
every sample equally, so the differences between consecutive samples stay
exactly as they were. Physics-consistent manipulation applies such an
offset, and change-point detection reads an AUC of $\pn{DerivativeBlindnessOffsetPlantTierOneAuc}{0.500}$, chance, on the
plant at all \pnc{CountTiers}{4} tiers and within $\pn{DerivativeBlindnessOffsetSimulatorEarlyTiersMaxDeparture}{0.001}$ of chance on the simulator at
tiers~1 through~3. At
simulator tier~4 it reads $\pn{DerivativeBlindnessOffsetSimulatorTierFourAuc}{0.445}$, a departure the offset argument does not
predict and we do not explain.

The second is saturation, on the plant only, since physics violation has no
simulator cells. Physics violation pins sensors to a fixed value,
and a pinned sensor stops changing altogether. It is therefore
\emph{calmer} than a healthy channel, which always carries some noise. The
detector ranks it as less suspicious than normal data, giving an AUC of
$\pn{DerivativeBlindnessPhysicsViolationChangePointMinAuc}{0.440}$ to $\pn{DerivativeBlindnessPhysicsViolationChangePointMaxAuc}{0.447}$: below chance, not at it.

This is specific to what the detector measures rather than a general
weakness. The same detector reaches $\pn{DerivativeBlindnessChangePointBestCellAuc}{1.000}$ on \pnc{DerivativeBlindnessChangePointCeilingDisturbances}{4} other disturbances.

On physics violation, change-point detection is the only one of \pnc{CountPanelDetectors}{11}
detectors below chance at every tier; NTS-NOTEARS stays above chance, rising
from $\pn{DerivativeBlindnessPhysicsViolationNtsNotearsTierOneAuc}{0.66}$ at tier~1 to $\pn{DerivativeBlindnessPhysicsViolationNtsNotearsTierFourAuc}{0.93}$ at tier~4, where change-point detection stays
flat. The level-based detectors are unaffected: PCA reconstruction reads
$\pn{DerivativeBlindnessPhysicsViolationPcaReconstructionTierFourAuc}{1.0000}$, Hotelling's $T^2$ $\pn{DerivativeBlindnessPhysicsViolationHotellingTSquaredTierFourAuc}{0.9918}$ and the autoencoder $\pn{DerivativeBlindnessPhysicsViolationAutoencoderTierFourAuc}{0.9652}$. A sensor frozen
far from its normal value is obvious to a detector that compares levels and
invisible to one that compares consecutive samples. Both gaps follow
from the scoring rule, apart from the simulator tier-4 departure above, and we
expect other detectors built on rates of change to share them, though we did not
test one.

Our cost comparison points away from structure learning, and the overall
result invites the conclusion that a few classical detectors are enough.
That conclusion is safe only if the detectors kept do not all measure the
same kind of quantity. A detection rate averaged over disturbances would
not show the difference, because the disturbance one detector cannot see
is one another catches perfectly.

\label{sec:transfer}

\begin{table}[t]
\caption{Rank correlation between the processes by operating point: $\rho$ is
Spearman's, $\tau_b$ Kendall's tie-corrected coefficient, and the interval a
percentile bootstrap over disturbances. $n$ is the detectors ranked, \pnc{TransferAucDetectorCount}{11}
under AUC and \pnc{TransferBudgetOnePermilleDetectorCount}{10} with a resolvable operating point below.}
\label{tab:transfer}
\centering
\begin{tabular}{@{}lrrrr@{}}
\toprule
Metric & $n$ & $\rho$ & $\tau_b$ & \pn{TransferKendallTauCiLevel}{95\%} CI on $\tau_b$ \\
\midrule
AUC                 & \pn{TransferAucDetectorCount}{11} & \pn{TransferAucSpearmanRho}{0.745} & \pn{TransferAucKendallTau}{0.564} & $[\pn{TransferAucKendallTauCiLow}{+0.16}, \pn{TransferAucKendallTauCiHigh}{+0.78}]$ \\
TPR at FPR $=\pn{BudgetOnePermille}{0.001}$ & \pn{TransferBudgetOnePermilleDetectorCount}{10} & \pn{TransferBudgetOnePermilleSpearmanRho}{0.650} & \pn{TransferBudgetOnePermilleKendallTau}{0.539} & $[\pn{TransferBudgetOnePermilleKendallTauCiLow}{+0.13}, \pn{TransferBudgetOnePermilleKendallTauCiHigh}{+0.67}]$ \\
TPR at FPR $=\pn{BudgetFivePermille}{0.005}$ & \pn{TransferBudgetFivePermilleDetectorCount}{10} & \pn{TransferBudgetFivePermilleSpearmanRho}{0.614} & \pn{TransferBudgetFivePermilleKendallTau}{0.494} & $[\pn{TransferBudgetFivePermilleKendallTauCiLow}{+0.09}, \pn{TransferBudgetFivePermilleKendallTauCiHigh}{+0.63}]$ \\
TPR at FPR $=\pn{BudgetOnePercent}{0.010}$ & \pn{TransferBudgetOnePercentDetectorCount}{10} & \pn{TransferBudgetOnePercentSpearmanRho}{0.614} & \pn{TransferBudgetOnePercentKendallTau}{0.494} & $[\pn{TransferBudgetOnePercentKendallTauCiLow}{+0.09}, \pn{TransferBudgetOnePercentKendallTauCiHigh}{+0.60}]$ \\
TPR at FPR $=\pn{BudgetTwoPercent}{0.020}$ & \pn{TransferBudgetTwoPercentDetectorCount}{10} & \pn{TransferBudgetTwoPercentSpearmanRho}{0.614} & \pn{TransferBudgetTwoPercentKendallTau}{0.494} & $[\pn{TransferBudgetTwoPercentKendallTauCiLow}{+0.09}, \pn{TransferBudgetTwoPercentKendallTauCiHigh}{+0.63}]$ \\
TPR at FPR $=\pn{BudgetFivePercent}{0.050}$ & \pn{TransferBudgetFivePercentDetectorCount}{10} & \pn{TransferBudgetFivePercentSpearmanRho}{0.340} & \pn{TransferBudgetFivePercentKendallTau}{0.315} & $[\pn{TransferBudgetFivePercentKendallTauCiLow}{-0.13}, \pn{TransferBudgetFivePercentKendallTauCiHigh}{+0.73}]$ \\
\bottomrule
\end{tabular}
\end{table}

\emph{Transfer of the ranking.} Table~\ref{tab:transfer} ranks the \pnc{TransferAucDetectorCount}{11} detectors best to worst on each
facility and compares the \lit{two} orderings, at every operating point and under
AUC. Spearman's $\rho$ and Kendall's $\tau_b$ are both $1$ if the orderings
agree exactly, $0$ if unrelated and negative if inverted. Both are reported
because at a tight budget most plant detectors sit at or near zero, and
$\rho$ overstates agreement on a largely tied vector. The $b$ in $\tau_b$ is
that correction: pairs tied on either process are excluded from the count
of agreements and discounted in the normalizing total, instead of being
credited as agreement. 

\lit{Two} questions follow: how uncertain those coefficients are, and whether they
could arise by chance. Each detector's entry is defined per facility as follows: within each of the
\pnc{TransferDisturbanceCount}{7} disturbances both processes share, the median over tiers~1 through~4
and over seeds of its rate at the operating point (or of its AUC), then the
median of those \pnc{TransferDisturbanceCount}{7} values. The ranking is therefore not the tier-2 ranking
of Table~\ref{tab:detection}. What would differ on a repeat of
this study is which disturbances were drawn, not which detectors. The
intervals therefore resample disturbances. A Fisher-z interval would instead
treat those medians as exact and the \pnc{TransferAucDetectorCount}{11} detectors as the sample,
describing uncertainty about a population of detectors we did not draw from.
Every interval excludes zero except the one at the $\pn{BudgetFivePercent}{0.050}$ budget, which
runs from $\pn{TransferBudgetFivePercentKendallTauCiLow}{-0.13}$ to $\pn{TransferBudgetFivePercentKendallTauCiHigh}{+0.73}$.

The \pnc{CountBudgets}{5} budgets read the same curves at \pnc{CountBudgets}{5} points, so they are not \pnc{CountBudgets}{5}
independent tests and are not assessed as \pnc{CountBudgets}{5}. Shuffling which detector is
which, with the same shuffle applied at every budget, gives an observed mean
$\tau_b$ of $\pn{TransferJointMeanKendallTau}{0.47}$ at $p\approx \pn{TransferJointPermutationP}{0.02}$, a mean this large
about one time in \pn{TransferJointPermutationOneInN}{41} under random relabeling. Its significance depends
on what is aggregated. Restricted to tier~2, the joint mean $\tau_b$ is
$\pn{TransferSensitivityTierTwoJointMeanKendallTau}{0.23}$ at $p\approx \pn{TransferSensitivityTierTwoJointPermutationP}{0.16}$. Pooling all disturbance-tier
cells instead of taking per-disturbance medians gives
$\pn{TransferSensitivityPooledJointMeanKendallTau}{0.26}$ at $p\approx \pn{TransferSensitivityPooledJointPermutationP}{0.15}$. Omitting one disturbance at a time gives $p$ from
$\pn{TransferSensitivityLeaveOneOutMinimumJointPermutationP}{0.03}$ to $\pn{TransferSensitivityLeaveOneOutMaximumJointPermutationP}{0.09}$, with the mean positive in every case
($\pn{TransferSensitivityLeaveOneOutPositiveCount}{7}$ of \pnc{TransferDisturbanceCount}{7}). The direction is robust and the significance is not:
we claim a positive association, not a significant one. The permutation tests random matching of detector labels, not
replication on another process pair. The bootstrap intervals above
resample disturbances rather than detector labels, so they answer a
different question. Computed separately for each of
the \pnc{TransferDisturbanceCount}{7} disturbances, $\tau_b$ is positive in all \pnc{TransferDisturbancePositiveCount}{7}.

Hartung et al.\ report
rankings that shift with both metric and scenario
\cite{hartungTennesseeEastmanBenchmarking2026}; we find a rank correlation
that stays positive across \pnc{CountProcesses}{2} processes at every operating point examined
and never a preserved
ordering: at the $\pn{BudgetOnePermille}{0.001}$ budget, with $\tau_b = \pn{TransferDiscordantKendallTau}{0.54}$, about $\pn{TransferDiscordantPairShare}{22\%}$ of
detector pairs are discordant. The \lit{two} results are therefore not in clear conflict. The designs
differ in a way that may account for the difference: their rankings move across
scenarios, each process contributing its own faults, whereas one
construction is applied to both processes here, so what varies is the
telemetry and not the disturbance. Comparing
detector orderings between \lit{two} disturbances within one process, at the same
budget and on the same per-disturbance basis as the cross-process figure,
the median $\tau_b$ is $\pn{TransferStabilityPlantMedianKendallTau}{0.422}$ on the plant and $\pn{TransferStabilitySimulatorMedianKendallTau}{0.484}$ on the simulator,
against $\pn{TransferStabilityCrossProcessMedianKendallTau}{0.598}$ across processes. A ranking is thus more stable under a
change of plant than under a change of what is being detected, which is the
reverse of the expected robustness order. A plausible explanation
consistent with this descriptive comparison, not an established mechanism, is
that their rankings moved because their scenarios did.

The ranking, then, carries over at most weakly even where absolute detection
rates are not preserved. Which detector is relatively stronger on the
other process is somewhat predictable from this one; the absolute rate it
reaches there, as the \lit{two} process results above show, is not preserved. Those are compatible,
and the second does not follow from the first: a panel can keep part of its
internal order across a move that lowers the detection rate of most members.

How far the estimates can be pushed is limited by the panel. \Pnc{TransferBudgetOnePermilleDetectorCount}{10} or \pnc{TransferAucDetectorCount}{11}
detectors is a small sample for a rank statistic, the intervals all overlap,
and the ordering across budgets carries no information: what the table
supports is that the point estimates are positive throughout, not any particular
value at any particular budget.

PCMCI+ is ranked under AUC and held out of the operating-point rows, because it
produces about \pnc{DesignPlantPcmciPlusMedianRocVertexCountRounded}{11} distinct scores across about \pn{DesignPlantPcmciPlusMedianNegativeCount}{11,503}
evaluation negatives: enough for a full curve, but leaving it essentially one usable vertex
at a tight budget, where its rank would report where those few steps happen to
fall. Figures~\ref{fig:budget} and~\ref{fig:heatmap} exclude it on the same
grounds. Including it raises $\rho$ at all \pnc{CountBudgets}{5} budgets, to $\pn{TransferSensitivityBudgetOnePermilleSpearmanRho}{0.729}$, $\pn{TransferSensitivityBudgetFivePermilleSpearmanRho}{0.702}$,
$\pn{TransferSensitivityBudgetOnePercentSpearmanRho}{0.702}$, $\pn{TransferSensitivityBudgetTwoPercentSpearmanRho}{0.702}$ and $\pn{TransferSensitivityBudgetFivePercentSpearmanRho}{0.487}$; every estimate stays positive, so the exclusion
changes the size of the correlation rather than its sign.

\label{sec:robustness}

\emph{Robustness checks.} \lit{Three} checks bear on whether the result above is an artifact of the
measurement rather than of the data: seed variation, a limit-alarm baseline, and the clamping mode. Each
reads the real facility's corpora.

\emph{Seed variation.} All \pnc{CountPanelDetectors}{11} detectors run at \pnc{DesignPlantSeedsPerDetectorCount}{5} seeds on both
processes. \Pnc{SeedSpreadIdenticalAucFourDecimalsDetectorCount}{6} of them (change-point detection, Hotelling's $T^2$, PCA reconstruction,
dependency violation, DYNOTEARS and PCMCI+) return the same AUC at every seed
to \lit{four} decimal places.
On the plant, the across-seed standard deviation of AUC, averaged over the
$\pn{SeedSpreadPlantDisturbanceTierCells}{36}$ disturbance-tier cells, is largest for NTS-NOTEARS at $\pn{SeedSpreadPlantNtsNotearsMeanAucSd}{0.071}$, reaching
$\pn{SeedSpreadPlantNtsNotearsMaxAucSd}{0.203}$ in its worst cell, and for the autoencoder at $\pn{SeedSpreadPlantAutoencoderMeanAucSd}{0.039}$, reaching
$\pn{SeedSpreadPlantAutoencoderMaxAucSd}{0.125}$. The other \pnc{SeedSpreadOtherVaryingDetectorCount}{3} that vary sit at or below $\pn{SeedSpreadOtherVaryingMeanAucSdMax}{0.019}$. The results of those \pnc{SeedSpreadUnstableDetectorCount}{2}
detectors should be read with that in mind in the tier ladder and in
Table~\ref{tab:detection}.

\Pnc{DesignPlantSeedsPerDetectorCount}{5} seeds can show that a detector varies; they cannot show that one does
not. A spread as large as NTS-NOTEARS's or the autoencoder's is hard to draw
by chance \pnc{DesignPlantSeedsPerDetectorCount}{5} times over, so their instability is established, and their
contribution to any result here could be an accident of seed. A small or zero
spread is consistent with a larger one that \pnc{DesignPlantSeedsPerDetectorCount}{5} draws happened to miss, so
the other \pnc{SeedSpreadLooselyBoundedDetectorCount}{9} are bounded loosely rather than shown to be stable.

Those figures are AUC, while results are reported at an operating point, so
we compare the \lit{two}. Over the $\pn{SeedSpreadOperatingCellCount}{216}$ cells of the \pnc{DesignRandomStepDetectorCount}{6} detectors with a random
step by design (autoencoder, LSTM, isolation forest, NTS-NOTEARS, DYNOTEARS and
PCMCI+; the other \pnc{DesignDeterministicDetectorCount}{5} have none), the operating point looks the more stable
(median across-seed standard deviation $\pn{SeedSpreadOperatingMedianTprSd}{0.0000}$ against $\pn{SeedSpreadOperatingMedianAucSd}{0.0089}$), but that is
a floor effect: $\pn{SeedSpreadOperatingAlwaysZeroCellCount}{63}$ cells return zero at every seed and a statistic pinned at
zero cannot vary. Over the $\pn{SeedSpreadOperatingDetectingCellCount}{71}$ cells that detect anything here, a median TPR
above $\pn{SeedSpreadOperatingDetectingMedianTprFloor}{0.05}$, the \lit{two} are close, $\pn{SeedSpreadOperatingDetectingMedianTprSd}{0.0047}$ for TPR against $\pn{SeedSpreadOperatingDetectingMedianAucSd}{0.0062}$ for AUC,
so the AUC check does not understate the operating-point spread where
detection is happening. A tail remains: $\pn{SeedSpreadOperatingUnstableCellCount}{31}$ cells have a TPR standard
deviation above $\pn{SeedSpreadOperatingTprSdThreshold}{0.060}$.
In Table~\ref{tab:detection}, \pn{DetectionSeedMarkCount}{6} entries are marked because leaving out any one seed moves
their median by more than $\pn{DetectionSeedMarkThreshold}{0.06}$, \pn{DetectionPlantSeedMarkCount}{4} on the plant and \pn{DetectionSimulatorSeedMarkCount}{2} on the
simulator; no other entry does. All of them belong to the neural and causal
families, so the instability qualifies the family comparison rather than
supporting it.

\emph{The limit alarm.} Table~\ref{tab:threat} expects a limit alarm to
catch the severe tiers and not the mild ones, which is why tier~2 is the tier
reported. Bounded mode clamps
values back inside the support the alarm reads, so the alarm is scored on the
unbounded corpus. Scored as a detector on the same per-sample basis as the
panel, its rate on physics violation is $\pn{LimitAlarmPhysicsViolationTierTwoTpr}{0.000}$ through tier~2 and rises to
$\pn{LimitAlarmPhysicsViolationTierThreeTpr}{0.667}$ and $\pn{LimitAlarmPhysicsViolationTierFourTpr}{0.833}$ at tiers~3 and~4, which is the behavior the table
describes. Across the \pnc{LimitAlarmBoundedDisturbances}{8} disturbances that perturb a bounded channel (operator mimicry
perturbs none, so the alarm has nothing to read there), the median
rate is $\pn{LimitAlarmTierThreeMedianTpr}{0.000}$ through tier~3 and reaches only $\pn{LimitAlarmTierFourMedianTpr}{0.109}$ at tier~4, so most disturbances at those tiers
are not caught cheaply, and the table says ``the disturbances that leave the
range'' rather than naming the tiers.

The alarm is quiet: its false-alarm rate is $\pn{LimitAlarmFalseAlarmRateMax}{0\%}$ in every cell, so it does
not buy detection with alarms. It also sees
almost nothing at the tiers the paper reports. Flow oscillation is flagged at
$\pn{LimitAlarmFlowOscillationTierFourTpr}{0.188}$ at tier~4 and at $\pn{LimitAlarmFlowOscillationTierThreeTpr}{0.000}$ below it. A cheap control that is quiet but does not see such
disturbances until they are large is not a substitute for monitoring. The
alarm also exists only for sensors, so an actuator-side disturbance is outside
what a limit check can look at, rather than something it looks at and misses.

\emph{Clamping mode.} Paired per-seed comparison of mean AUC between the \pnc{DesignPlantClampingModeCount}{2}
modes of the tiering rules puts the largest per-detector difference at
$\pn{ClampModeMaxMeanAucDifference}{0.010}$, with none reaching significance, though at $n=\pn{ClampModePairedSeedCount}{5}$ the smallest
attainable \lit{two}-sided Wilcoxon $p$ is $\pn{ClampModeMinimumTwoSidedP}{0.0625}$, so this is a weak null and not
a demonstration of equivalence.

\section[Discussion]{Discussion\aimark}\label{sec:discussion}

Detector performance measured on a simulated facility should not be taken at
face value as evidence of performance on a real one. That was already
established for native process faults scored by F1 and precision--recall area
\cite{hartungTennesseeEastmanBenchmarking2026}. The failure is not an
artifact of those choices: it persists for adversarial disturbances, under
one construction applied to both processes, and at a declared false-alarm
budget.

Cost bears on this at family level, the only level at which our measurement
reproduced. In \lit{two} September runs that scored one cell with the training cache
disabled, the \pnc{CountCausalFamilyDetectors}{3} causal detectors took the \pnc{CountCausalFamilyDetectors}{3} longest to score the cell (the
scoring phase alone, excluding training) of the detectors that timed
successfully, with no other detector inside their range, and the neural
detectors' scoring times bracketed the classical ones rather than separating
from them. Training time orders the detectors differently and is not compared. The
change-point detector's timing failed in both runs, so it is missing from that
comparison. The objection that inference cost rules out learned detectors
therefore does not select against the neural family here; it selects against
structure learning.

It would be a misreading to take this as an argument that statistical methods
win. They won here, on \pnc{CountProcesses}{2} processes held at steady state, against
disturbances that are perturbations of a stationary operating point. The
returns on model capacity are far easier to demonstrate on a simulator than on
a plant, while the costs enumerated in Section~\ref{sec:intro} (qualification
burden, inference budget, labeled-data scarcity, and requalification at each
new facility) all fall due on the plant. On this evidence, performance on the simulated
process does not reproduce on the real one under one construction.

The conditions we would most expect to reverse the conclusion are absent from
this record, and absent by what the processes did rather than by our choice. As Section~\ref{sec:design} reports, $\pn{DesignRegimeSteadyStateSegmentShare}{96.4\%}$ of
the plant's labeled segments are steady state and none at all are startup,
shutdown or trouble. A process in transition (startup, shutdown, a grade or
campaign change, multi-mode operation) breaks the stationarity that makes a
fixed projection and a residual threshold adequate, and there the ability to
learn a nonlinear, regime-dependent model of normal operation stops being
surplus capacity and becomes the requirement. We would expect the ordering
reported here to invert. Nothing in our design tests it, because the record
contains almost none of it. For a plant that spends much of its life in
transition this paper is out of scope rather than contrary evidence.

\section[Limitations and Threats to Validity]{Limitations and Threats to Validity\aimark}\label{sec:limitations}

The simulator's bounds are mostly empirical. \Pnc{SimulatorBoundsDeclaredChannels}{2} channels, reactor pressure
and reactor temperature, take value limits from the process description; the
other \pnc{SimulatorBoundsObservedRangeChannels}{51} use their own observed range with padding, and the
per-channel rate ceilings sit at twice each channel's observed maximum rate,
an auditable heuristic rather than a physics derivation. Its tier machinery
is therefore calibrated against what that process has been seen to do, and
the claim that its severest tier violates the process's physics is not
supported. The plant declares no per-channel bounds, so its
envelope is derived from its own telemetry by heuristic as well.

Disturbances are synthetic on both processes, which is what makes the
comparison controlled, one construction applied to both with magnitudes drawn
from each process's own admissible support; it is not a claim about
disturbances observed in the wild.

Nor is any of them adaptive. Each is derived from a control flaw and drawn
against an admissibility envelope; none is optimized against a detector, and
no attack here responds to what is deployed. An adversary who can query a
deployed detector, or who maximizes process impact subject to staying under
its threshold \cite{urbinaLimitingImpactStealthy2016}, is a stronger threat
than anything measured here; concealment attacks constructed against
reconstruction-based detectors are the concrete demonstration
\cite{erbaAssessingModelfreeAnomaly2022}. The rates in
Section~\ref{sec:results} describe non-adaptive disturbances only and are not
estimates of what such an adversary would achieve.

Actuator-side disturbances are injected open loop. Values are written into
recorded telemetry after the fact, so a controller that would have responded
to a valve command or a setpoint change does not, and the downstream
channels carry no compensating motion. Whether that makes such a disturbance
easier or harder to detect is not determined: the absence of a response
removes a cue, and the absence of compensation leaves the primary deviation
larger. The simulator could be driven closed loop and was not.

Holding a configuration constant across processes is not the same as holding
it neutral. The top-$k$ aggregation described in the supplementary appendix is
the clearest case: a fixed $k$ is a larger share of the simulator's channels
than of the plant's, so the same nominal setting is more selective on the
plant. Varying it per process would have put a tuning difference into a
comparison premised on identical configuration, so it was left alone.

Thresholds are set on data the detectors were trained on, so the
false-positive rate is estimated on scores the model has already seen. The
bias this introduces across detector families and across the \pnc{CountProcesses}{2} processes is
unmeasured. A held-out block would measure it: blocked time-ordered
cross-validation inside the training span is standard, and was not run for
compute reasons.

Each detector also runs at a single tuning budget, and \pnc{DesignChannelCappedDetectorCount}{4} at a reduced
channel budget, whose effect on detection was not measured.

\section[Conclusion]{Conclusion\aimark}\label{sec:conclusion}

On the \pnc{CountProcesses}{2} processes studied here, detection performance on a
simulated process did not reproduce on a real one. A similar gap was already reported
for native process faults scored without a fixed false-alarm rate; it appears under the \lit{three}
changes a security evaluation makes, to attacker-authored disturbances, one
construction across both processes, and a declared false-alarm budget.

The scope is narrow. \Pnc{CountProcesses}{2} processes held almost entirely at steady state, the
plant by its regime labels and the simulator by construction, and no account of what
causes the lower detection on the plant. A plant that spends much of its life in
transition is a different question, and one with more room for learned
capacity to earn its inference and requalification cost.

The ranking carries over, though not as an ordering. Which detector is relatively
stronger is positively correlated across the \pnc{CountProcesses}{2} processes at every operating
point examined; absolute detection rates are not preserved and differ
unevenly across detectors and disturbances, the neural family lowest on the plant.

\section*{Acknowledgment}

Generative AI was used in preparing every section of this article, including
its text, figures, tables and analysis code. Anthropic Claude~\cite{aiClaude}
wrote and revised the analysis software and drafted and edited text; OpenAI
GPT~\cite{aiGPT} and Google Gemini~\cite{aiGemini} reviewed the code and
manuscript; and a literature-screening tool built on those models and on
Qwen~\cite{aiQwen}, Granite~\cite{aiGranite}, Llama~\cite{aiLlama} and
BGE~\cite{aiBGE} searched and assessed related work. The author reviewed, verified and revised all AI-generated content and
takes full responsibility for it. Every quantitative result was regenerated
by running the analysis code rather than taken from model output.

\bibliographystyle{IEEEtran}
\bibliography{local,../../sources}

\begin{IEEEbiographynophoto}{Benjamin Blakely}
(Senior Member, IEEE) is a cybersecurity and artificial intelligence
researcher at Argonne National Laboratory, where he leads the Applied
Research group within the Strategic Security Sciences division. His work
centers on how AI and control systems behave under adversarial conditions,
and what it takes to trust automated decision-making for critical
infrastructure protection. Over 20 years across industry, government, and
academia he has built and led teams in information security and data
science, and he serves as a subject matter expert for United States
departments and agencies including Energy, NNSA, Homeland Security, State,
and Defense. He holds PhD and BS degrees in computer engineering from Iowa
State University and is a CISSP through ISC\textsuperscript{2}.
\end{IEEEbiographynophoto}

\section*{Supplementary Appendix: Detector Configuration}
\setcounter{table}{0}
\renewcommand{\thetable}{A.\Roman{table}}
\noindent Supplement to \emph{Adversarial Process Disturbances at a Deployable
Alarm Budget: Detector Evaluation on a Simulator and a Real Plant}.
\emph{AI-assisted content:} this material was prepared with generative AI ---
Anthropic Claude, OpenAI GPT, Google Gemini, Alibaba Qwen, IBM
Granite, Meta Llama and BAAI BGE models --- reviewed and verified by the
author. The article's Acknowledgment gives the role of each system and the
full citations.

\section*{Detector Configuration}

Table~\ref{tab:hyper} gives one configuration per detector, used for every
cell and fixed before the comparison. That is what makes the panel
comparable, and also a limitation: a detector that would do better with
process-specific tuning is reported without it.

\begin{table}[h]
\caption{Detector configuration, identical across both processes; top-$k$ $30$
and PCMCI+'s lag $2$ come from sweeps, the channel ceilings are compute
budgets, and library defaults fill the estimator count, explained-variance
fraction, $\alpha$ and selection criterion. Everything else, including every
neural hyperparameter, was set once and never swept.}
\label{tab:hyper}
\centering
\footnotesize
\begin{tabular}{@{}lp{5.2cm}@{}}
\toprule
Detector & Configuration \\
\midrule
Hotelling's $T^2$ & window $\pn{HyperHotellingWindow}{500}$, stride $\pn{HyperHotellingStride}{250}$, threshold pct.\ $\pn{HyperHotellingThresholdPercentile}{95}$ \\
PCA reconstruction & explained variance $\pn{HyperPcaExplainedVariance}{0.95}$, top-$k$ $\pn{HyperPcaTopK}{30}$, window $\pn{HyperPcaWindow}{500}$, stride $\pn{HyperPcaStride}{250}$, threshold pct.\ $\pn{HyperPcaThresholdPercentile}{85}$ \\
Isolation forest & $\pn{HyperIsolationForestEstimators}{100}$ estimators, max samples auto, window $\pn{HyperIsolationForestWindow}{500}$, stride $\pn{HyperIsolationForestStride}{250}$ \\
Change-point detection & weighted first difference, second difference and rolling variance ($\pn{HyperChangePointFirstDifferenceWeight}{0.5}/\pn{HyperChangePointSecondDifferenceWeight}{0.3}/\pn{HyperChangePointRollingVarianceWeight}{0.2}$), top-$k$ $\pn{HyperChangePointTopK}{30}$ over sensors, threshold pct.\ $\pn{HyperChangePointThresholdPercentile}{85}$ \\
Autoencoder & latent $\pn{HyperAutoencoderLatent}{4}$, dropout $\pn{HyperAutoencoderDropout}{0.1}$, batch norm, lr $10^{-\pn{HyperAutoencoderLearningRateExponent}{3}}$, batch $\pn{HyperAutoencoderBatch}{32}$, $\pn{HyperAutoencoderEpochs}{20}$ epochs, top-$k$ $\pn{HyperAutoencoderTopK}{30}$ \\
LSTM & hidden $\pn{HyperLstmHidden}{40}$, dropout $\pn{HyperLstmDropout}{0.25}$, lr $10^{-\pn{HyperLstmLearningRateExponent}{3}}$, batch $\pn{HyperLstmBatch}{16}$, $\pn{HyperLstmEpochs}{20}$ epochs, patience $\pn{HyperLstmPatience}{8}$, sequence $\pn{HyperLstmSequence}{50}$, top-$k$ $\pn{HyperLstmTopK}{30}$ \\
Dependency violation & correlation threshold $\pn{HyperDependencyViolationCorrelationThreshold}{0.3}$, Granger lags $\pn{HyperDependencyViolationGrangerMinimumLag}{1}$--$\pn{HyperDependencyViolationGrangerMaximumLag}{10}$, $\le \pn{HyperDependencyViolationGrangerMaximumPairs}{100}$ pairs \\
Bayesian network & $\le \pn{HyperBayesianNetworkMaximumChannels}{25}$ sensors, $\le \pn{HyperBayesianNetworkMaximumParents}{3}$ parents, Bayesian information criterion, window $\pn{HyperBayesianNetworkWindow}{500}$ \\
DYNOTEARS & $\le \pn{HyperDynotearsMaximumChannels}{30}$ channels, lag $\pn{HyperDynotearsLag}{3}$, edge threshold $\pn{HyperDynotearsEdgeThreshold}{0.01}$, $\lambda_W=\pn{HyperDynotearsLambdaW}{0.05}$, $\lambda_A=\pn{HyperDynotearsLambdaA}{0.05}$ \\
NTS-NOTEARS & $\le \pn{HyperNtsNotearsMaximumChannels}{30}$ channels, lag $\pn{HyperNtsNotearsLag}{3}$, hidden $\pn{HyperNtsNotearsHidden}{16}$, kernel $\pn{HyperNtsNotearsKernel}{3}$, $\lambda_W=\pn{HyperNtsNotearsLambdaW}{0.05}$, $\lambda_A=\pn{HyperNtsNotearsLambdaA}{0.05}$ \\
PCMCI+ & $\le \pn{HyperPcmciMaximumChannels}{20}$ channels, lag $\pn{HyperPcmciLag}{2}$, $\alpha=\pn{HyperPcmciAlpha}{0.05}$, partial-correlation independence test \\
\bottomrule
\end{tabular}
\end{table}

\lit{Four} detectors produce one score per sensor per timepoint and must reduce
that to a single score, and how they reduce it affects what they can
detect. The autoencoder, LSTM, PCA reconstruction and change-point
detection all take the mean of the $30$ largest per-sensor scores at each
timepoint. The first \lit{three} standardize each sensor against its own baseline
median and absolute deviation before selecting, so a noisy channel cannot
dominate on scale alone; change-point detection selects on its raw scores.
Selection is redone at every timepoint, so it is not a fixed subset.

The value $30$ comes from a sweep over $k \in \{3, 10, 20, 30, 50\}$ on the
plant, in which $k=10$ and $k=20$ were worse than either end of the range.
Change-point detection uses the same $k$ as the other \lit{three}; reducing by the
maximum, which is $k=1$, would let a disturbance missing whichever channel
happened to be loudest contribute nothing to its score. One consequence carries into the cross-process comparison: a
fixed $k$ is a different share of each channel set, about a fifth of the
plant's $143$ against more than half of the simulator's $53$.

The channel caps carried by \pnc{DesignChannelCappedDetectorCount}{4} detectors are a separate restriction, on
what the model sees rather than on how per-sensor scores are combined;
Section~IV of the article describes them.
\end{document}